\documentclass[showpacs,prb,aps,twocolumn]{revtex4}
\usepackage{graphicx}
\usepackage{color}
\usepackage{dcolumn}
\usepackage{bm}
\usepackage{amsmath}

\def\gapx{\lower 2pt \hbox{$\buildrel>\over{\scriptstyle{\sim}}$\ }}
\def\lapx{\lower 2pt \hbox{$\buildrel<\over{\scriptstyle{\sim}}$\ }}
\def\j1j2{$J_1$$-$$J_2$}
\begin{document}
\title{Ground-state phase diagram of the quantum $J_1-J_2$  model on the honeycomb lattice}

\author{Fabio Mezzacapo$^1$}
\author{Massimo Boninsegni$^2$}

\affiliation{$^1$Max-Planck-Institut f\"ur Quantenoptik, Hans-Kopfermann-Str.1, D-85748, Garching, Germany\\
$^2$Department of Physics, University of Alberta, Edmonton, Alberta, Canada T6G 2E1 }
\date{\today}

\begin{abstract}
We study the ground-state phase diagram of the quantum $J_1-J_2$ model on the honeycomb lattice by means of an entangled-plaquette variational ansatz. Values of energy and relevant  order parameters are computed in the range $0\le {J_2}/{J_1} \le 1$. The system displays  classical order  for ${J_2}/{J_1}\lesssim 0.2$ (N\'eel), and for $J_2/J_1 \gtrsim 0.4$ (collinear). In the intermediate region,  the ground-state is disordered.  Our results show that the reduction of the half-filled Hubbard model to the model studied here does not yield  accurate predictions.
 \end{abstract}


\maketitle

Frustrated quantum antiferromagnets  are  a subject of current intense research, whose study is rendered even more timely by recent, intriguing proposals of  possible experimental realizations with ultracold atoms.\cite{cold}  Frustration can arise either from the geometry of the system, or  from competing  interactions, and can lead to the stabilization of novel, exotic (magnetic and non-magnetic)  phases of matter. 

A prototypical spin model,  featuring interaction-induced frustration, is the  spin-1/2 antiferromagnetic (AF) Heisenberg Hamiltonian in presence of next-nearest-neighbor (NNN) coupling (usually referred to as the \j1j2 model):
\begin{equation}
J_1 \sum_{\langle i,j \rangle}\mathbf{S}_i \cdot \mathbf{S}_j +J_2\sum_{\langle\langle i,j \rangle\rangle}\mathbf{S}_i \cdot \mathbf{S}_j 
\label{eq:ham}
\end{equation}
where ${\bf S}_i$ is a spin-1/2 operator associated to the $i$th lattice site and the first (second) summation runs over NN (NNN) sites. Periodic boundary conditions (PBC) are assumed.\cite{foot}

Model (\ref{eq:ham})  has been extensively studied on the square lattice,\cite{SQ} and has more recently elicited great interest on the honeycomb one,\cite{honey1, honey2, honey3, honey4, honey5, honey6, honey7, ran, schmidt}. For  $J_2$\ =\ 0 (i.e., with no  NNN interaction), the ground state (GS) of (\ref{eq:ham}) features AF long-range (N\'eel) order on both these two bipartite lattices.\cite{Sand} 
\begin{figure}[t] 
 \begin{center}
{\includegraphics[scale=0.36]{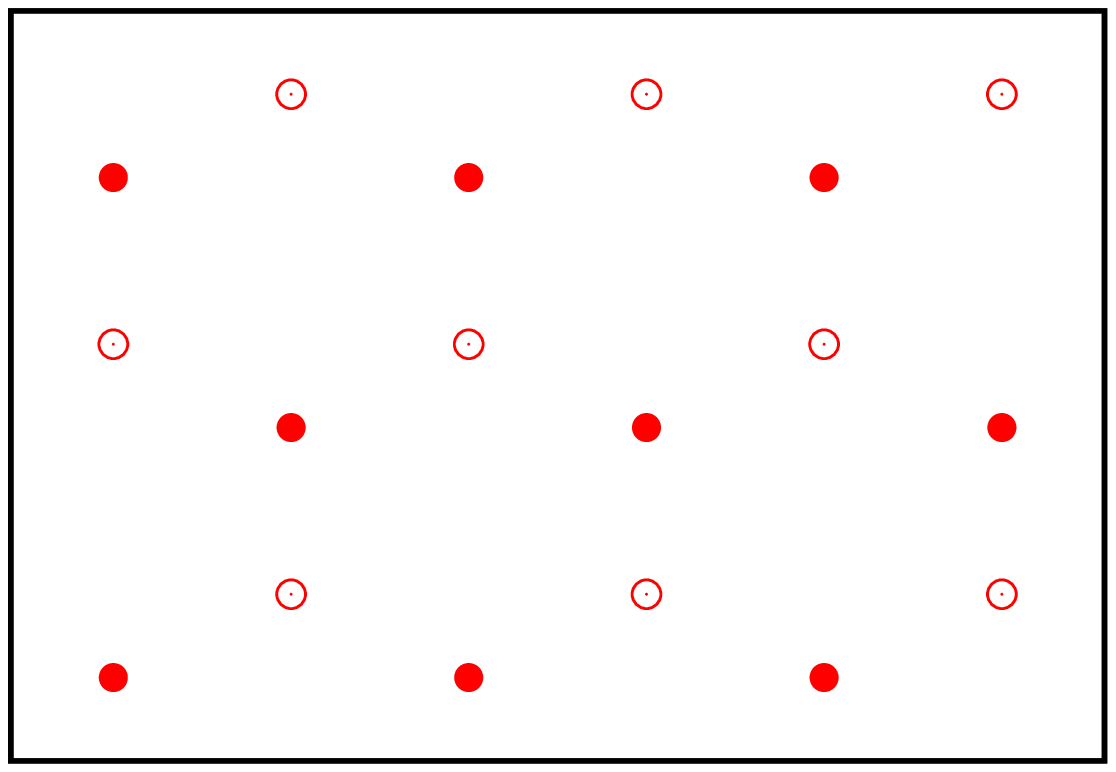}}
{\includegraphics[scale=0.36]{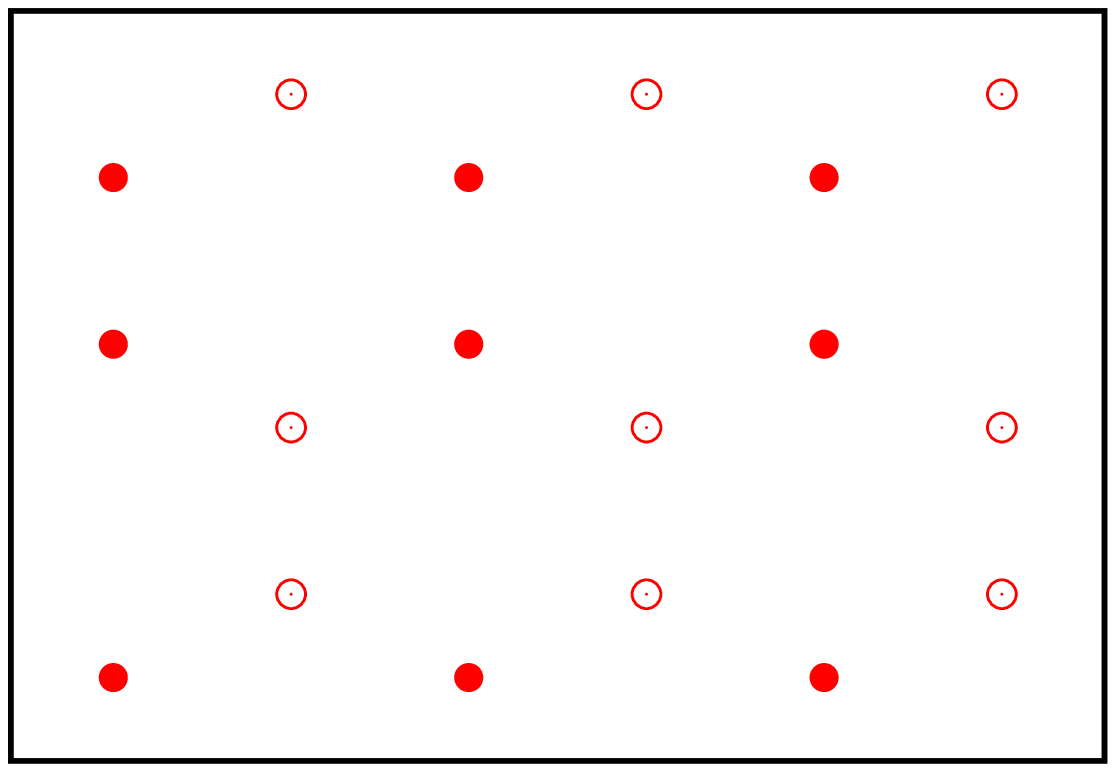}}
\end{center}
\caption{(color on line). Schematic representation of the honeycomb lattice, with the two different definitions of sublattices ($A$ and $B$) corresponding to N\'eel (left) and collinear (right) orders.}  
\label{fig:1}
\end{figure}
When $J_2 > 0$ the  system is frustrated. N\'eel order remains, for sufficiently small values of $J_2$, but as $J_2$ grows different phases, including disordered ones, become energetically competitive. For example, on the square lattice the GS displays magnetic long-range order (albeit of different types) for small and large value of $J_2$, while the nature of the GS in the intermediate region (i.e.,   ${J_2} \sim 0.5$)  is  still under debate.\cite{SQ} On the honeycomb lattice, due to its lower coordination number with respect to the square one, the disruptive effect  of quantum fluctuations on magnetic order is enhanced, as confirmed by the value of the sublattice magnetization, which  in the unfrustrated case is approximatively  10\% smaller on the honeycomb than on the square lattice \cite{Sand}.

Studies of the magnetic phase diagram of the \j1j2 model on the honeycomb lattice, based on different computational approaches, have yielded conflicting physical scenarios. For example, it is unclear if the GS of the system is disordered (``spin liquid") for any value of $J_2$, and even the nature of ordered phases remains controversial.\cite{honey1, honey2, honey3, honey4, honey5, honey6, honey7} Exact diagonalization  (ED)  of (\ref{eq:ham}) on small lattices (up to 42 sites),  yields evidence of AF N\'eel order for ${J_2}\lesssim 0.2$. For $0.2 \lesssim {J_2} \lesssim 0.4$ the system is predicted to be  in a plaquette valence-bond crystalline phase, while, for larger values of ${J_2}$, a collinear phase (not easy to characterize on small lattices) has been suggested \cite{honey3} (see Fig. \ref{fig:1}). The physical picture emerging  from  ED, whose predictive power is clearly affected by finite-size limitations,  must be assessed by calculations performed on systems of much larger  size, in order to carry out a reliable extrapolation of the relevant physical properties to the thermodynamic limit.  
 Series expansions yield, for instance, a similar physical scenario,  the main difference being a GS with spiral magnetic order  found in the intermediate region.\cite{honey7}

Quantum Monte Carlo (QMC) approaches yield numerically exact estimates for very large system sizes at  $J_2=0$, but are not applicable in presence of frustration due to the well-known ``sign" problem.  Conversely, variational Monte Carlo (VMC) methods are by definition free of such problem, and suffer from no significant size limitation. Of course, they are also approximate, their accuracy depending on the choice of the  variational wave function (WF). A recent variational investigation of the \j1j2 Hamiltonian on the honeycomb lattice, predicts the disappearance of N\'eel order at ${J_2} \simeq 0.08$, a  dimerized rotational symmetry-breaking phase for ${J_2} \gtrsim 0.3$ and a spin-liquid  GS in the intermediate region.\cite{honey1} It should be noted that, while in studies based on ED the presence of order of different kinds is assessed by a computation of relevant order parameters, the conclusions of Ref. \onlinecite{honey1} are mainly based on the comparison of energy estimates  yielded by  variational WF's corresponding to differently ordered, or disordered GS's. \\ \indent
Given the qualitative disagreement between ED and the only existing variational study, two basic aspects that still need be clarified are {\it a}) the type of magnetic order (if any)  of phase(s) that occur as  N\'eel order is suppressed on increasing $J_2$, and {\it b}) the nature of the GS for $J_2 \gtrsim 0.4$. Also of interest is the value of $J_2$ at which the N\'eel order vanishes,  which according to Ref. \onlinecite{honey1} is in quantitative agreement with that found for the half-filled Hubbard model.\cite{mura} 
\\ \indent
In this paper, we report results of  a variational study of the GS phase diagram of (\ref{eq:ham}) on the honeycomb lattice, using the variational family of entangled-plaquette states (EPS).\cite{epsor} This is a general ansatz,  which has been shown to yield accurate estimates of GS observables for quantum spin lattice Hamiltonians, including frustrated ones.\cite{epsor, epscps} The goal of this work is to determine the phase diagram of the model of our interest by direct estimation of the values of various order parameters, extrapolated to the thermodynamic limit. Specifically, besides GS energies we also compute sublattice magnetizations and dimer-dimer correlation functions.
We find AF N\'eel order up to $ J_2\lesssim 0.2$, whereas for $J_2 \gtrsim 0.4$ the AF order is collinear;  for $0.2 \lesssim {J_2} \lesssim 0.4$ the GS  is instead disordered. 

An exhaustive description of the EPS class of states can be found in Refs. \onlinecite {epsor} and \onlinecite{epscps}.  We merely wish to stress that the EPS WF  has been applied to a variety of lattice spin Hamiltonians, always providing results of accuracy at least comparable to (or better than) that afforded by different numerical techniques or variational WF's.

In this study, given the specific magnetic orders (see Fig. \ref{fig:1}) that we aim at identifying,  the appropriate phase factors  have been incorporated in the general  ansatz. Our numerical calculations have been carried out for lattices of size as large as $N=324$ sites. For each lattice size and value of ${J_2}$, we independently and systematically optimize the EPS WF, by progressively increasing the size of our plaquettes (i.e., the number $l$ of lattice sites that a plaquette comprises). Both the WF optimization and the evaluation of various physical quantities are achieved via the variational Monte Carlo method. We  extrapolate to the thermodynamic limit the estimates obtained on finite lattices of different sizes, for a given plaquette size, and then, asses the dependence of our  results on the plaquette size.
\begin{table}[t]
\caption{\label{tab:1}GS energy per site as a function of $J_2$ for a system of size $N=100$ with PBC. Data for different plaquette-size $l$ are shown. Variational estimates from Ref. \onlinecite{honey1} are also reported for comparison. Statistical errors (in parentheses) are on the last digit.}
\begin{ruledtabular}
\begin{tabular}{ccccc}
$J_2$ & $l=9$ & $l=16$ &$l=18$ & Ref. \onlinecite{honey1}  \\
\hline
0.0 & -0.54155(7) & -0.54418(6) & & -0.537901(9)  \\
0.1 & -0.49049(5) & -0.49349(6) & & -0.488209(3) \\
0.15 & -0.46744(5) &-0.47087(4) &  & -0.469448(2) \\
0.2 & -0.44599(4) &  -0.45004(6)  &-0.45145(4) & -0.450687(4) \\
0.3 & -0.40936(6) & -0.41580(7)  & -0.41751(4) & -0.41688(1) \\
0.4 & -0.40650(5) & -0.41185(5) & & -0.41125(1) \\
0.5 & -0.42432(5) & -0.42940(6)  & & -0.42808(2) \\
0.6 & -0.45248(5) &  -0.45712(4) & &   \\
0.7 & -0.48519(4)  &  -0.48989(5) & & -0.47377(3) \\
0.8 & -0.52112(6)  & -0.52610(5)  & & \\
0.9 & -0.55928(4) & -0.56489(5)  & & \\
1.0 & -0.59917(5) & -0.60534(5)  & & \\
\end{tabular}
\end{ruledtabular}
\end{table}
\begin{figure}[t]
{\includegraphics[scale=0.68]{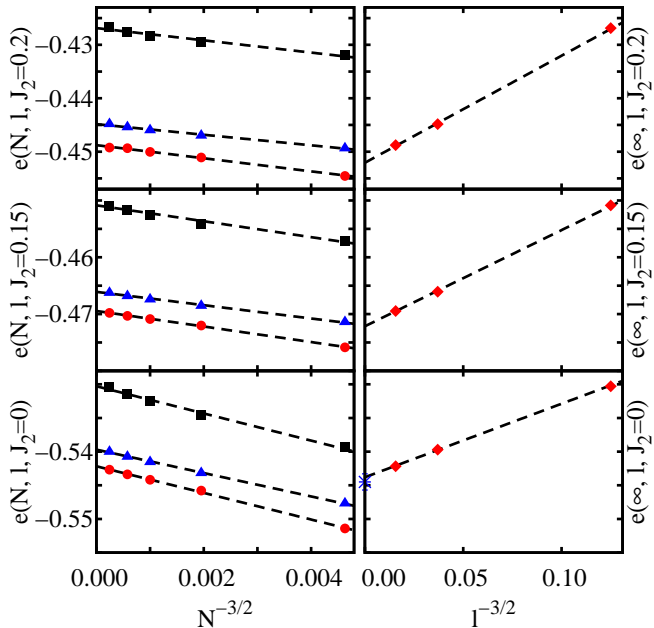}}
\caption{(color on line). Left: GS energy per site of the $J_1-J_2$ model on the honeycomb lattice with PBC, as a function of the system size. Estimates are obtained with $N$ plaquettes of $l=4$ (boxes), $l=9$ (triangles) and $l=16$ (circles) sites. Each dashed line is a fit  to numerical data with the same $l$ value. Right: GS energy per site, extrapolated to the thermodynamic limit, as a function of the plaquette-size. Each dashed line is a fitting function linear  in $l^{-3/2}$ (see text). The QMC estimate in the $J_2=0$ case is also shown for comparison (star).\cite{riera}}  
\label{fig:2}
\end{figure}

Table \ref{tab:1} shows the GS energy per site $e\equiv E/N$ for a lattice comprising $N=100$ sites,  for different values of $J_2$; estimates are shown corresponding to different plaquette sizes. Also shown for comparison are the variational estimates reported in  Ref. \onlinecite{honey1} by Clark {\it et al.} For all values of $J_2$, our energy estimates are lower than those of Ref. \onlinecite{honey1}. For $J_2 \le 0.1$, as well as for large $J_2$,  we obtain lower variational estimates than those of Ref. \onlinecite{honey1} already with plaquettes of size $l$=9. Plaquettes of larger size, typically  $l=16$ (and up to $l=18$ for values of $0.2 \lesssim J_2\lesssim0.3$),    are needed in order to improve on  the variational energy estimates of Ref. \onlinecite{honey1} in the remaining cases. This illustrates the main quality of the EPS ansatz, namely that it can be systematically improved by increasing the size of the plaquette, yielding energy estimates of  accuracy, typically higher than that afforded by other trial WF's.
\\ \indent
Estimates of the GS energy per site for various values of $J_2$ and different plaquette size $l$ are shown in Fig. \ref{fig:2} (left part), as a function of the system size. Extrapolation to the thermodynamic limit of  numerical estimates obtained with the same $l$ has been performed by assuming the functional scaling  form $e(N, l, J_2)=e(\infty, l, J_2)+\alpha(l, J_2)N^{-3/2}$.
Extrapolated values, therefore, depend on the size $l$ of the plaquette utilized. Since by construction expectation values (of {\it any} observable) must approach the {\it exact} GS values in the limit of large plaquette size, the question is how to obtain a reliable estimate for such a limit, based on those obtained with relatively small plaquettes. 
While in principle calculations with ever increasing plaquette size should be carried out, in order to observe numerical convergence of results, currently available computer resources limit the largest plaquette size for which this procedure can be implemented in practice. As it turns out, however, results for both the energy and for the relevant magnetic order parameters obtained with plaquettes of size up to $l$=16, lend themselves to simple numerical extrapolations.\cite{noteA} 
\\ \indent
An example of this procedure is illustrated in the right part of Fig. \ref{fig:2}, showing energy estimates extrapolated to the thermodynamic limit i.e., $e(\infty, l, J_2)$, as a function of the plaquette size. We assume that $e(\infty, l, J_2)$ can be expanded in powers of $l^{-1/2}$ and  fit the data  using the smallest number of powers. As shown in the figure, the values of $e(\infty, l, J_2)$ fall on a straight line when plotted versus $l^{-3/2}$, i.e., an excellent fit to the data is obtained with a single power (dashed lines); in any case, the extrapolated value does not change appreciably on including additional terms. For $J_2=0$, this procedure leads to an estimate  in agreement with that  obtained by QMC \cite{Sand, riera} (essentially exact in this case).

The cogent observable, in order  to assess the presence of magnetic order in the \j1j2 Hamiltonian, is the  squared sublattice magnetization (SSM), defined as
\begin{equation}
\mathbf{m}^2(N)=\biggl \langle\frac{1}{N^2}(\sum_{i\in A}\mathbf{S}_i-\sum_{j\in B}\mathbf{S}_j)^2\biggr \rangle
\end{equation}
where the two summations are taken over lattice sites belonging to different sublattices, as shown in Fig. \ref{fig:1}. The dependence of the values of the above observable on ${J_2}$, allows one to identify any range of parameters in which possibly magnetically disordered phases may exist. It is worth restating that in this work we consider only two types of magnetic long-range order, schematically illustrated in Fig. \ref{fig:1}, namely N\'eel and collinear.  
\begin{figure}[t] 
{\includegraphics[scale=0.68]{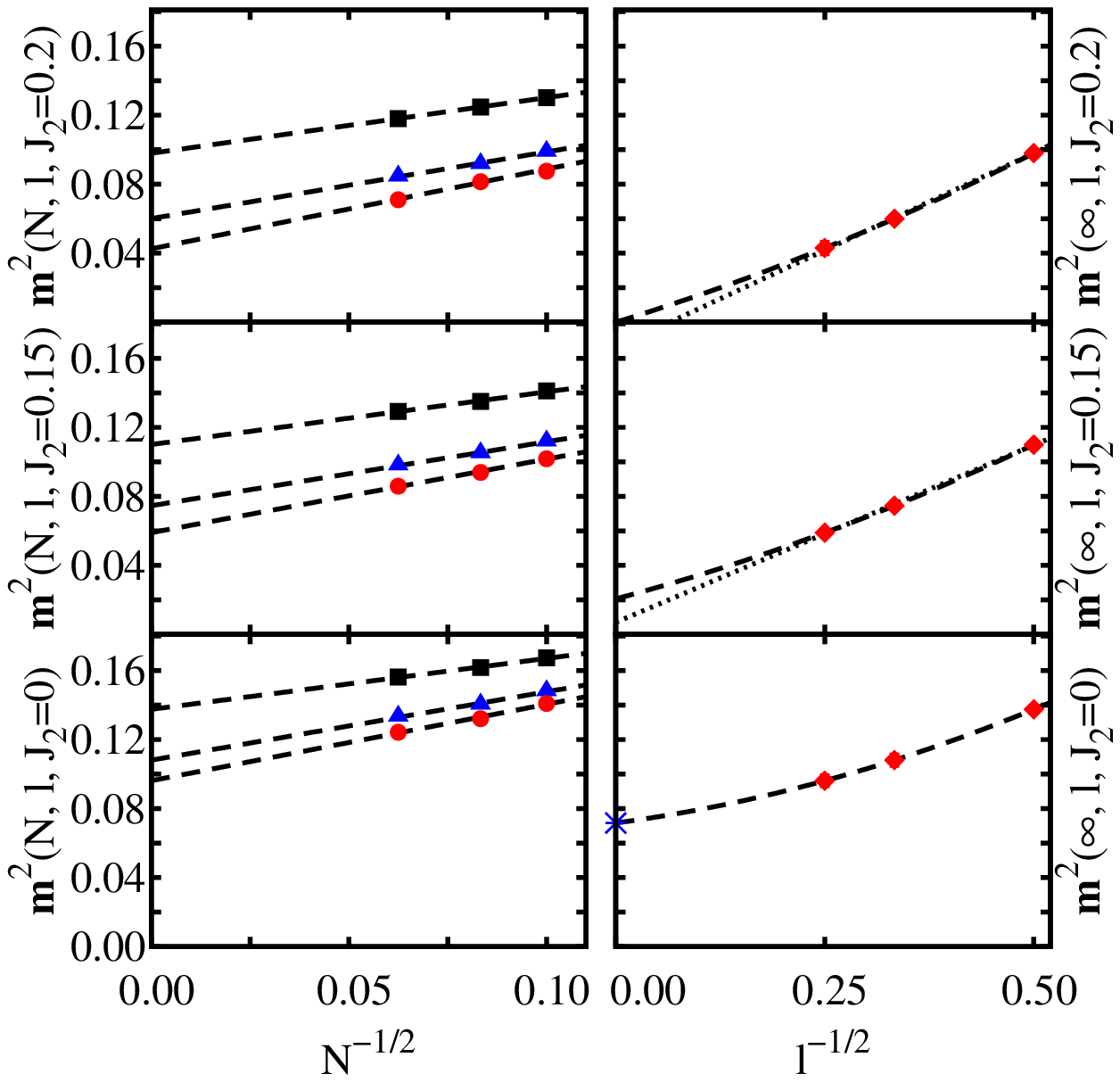}}
\caption{(color on line). Left: N\'eel  SSM of the $J_1-J_2$ model on the honeycomb lattice with PBC,  as a function of the system size. Estimates are obtained with $N$ plaquettes of $l=4$ (boxes), $l=9$ (triangles) and $l=16$ (circles) sites. Each dashed line is a fit  to numerical data with the same $l$ value (see text). Right: N\'eel  SSM, extrapolated to the thermodynamic limit, as a function of the plaquette-size. Lines are functions bult to infer the $l$ dependence of our data (see text). The QMC estimate in the $J_2=0$ case is also shown for comparison (star).\cite{Sand}}
\label{fig:3}
\end{figure}

Figure \ref{fig:3} shows estimates of the N\'eel SSM extrapolated to the thermodynamic limit  (left part), based on the functional form $\mathbf{m}^2(N, l, J_2)=\mathbf{m}^2(\infty, l, J_2)+\beta(l, J_2)N^{-1/2}$, for different plaquette sizes. Just like for the energy, the right part of the figure shows the extrapolated values $\mathbf{m}^2(\infty, l, J_2)$, this time plotted versus $l^{-1/2}$. The  fit to the data is performed using a quadratic functional dependence (dashed lines). 
\\ \indent  
For $J_2=0$, our  estimates in the large $l$ limit is again in quantitative agreement with QMC ones \cite{Sand, riera}. At $J_2=0.15$ the N\'eel SSM is finite; we make such a statement based on the observation that assuming a linear dependence of $\mathbf{m}^2(\infty, l, J_2)$ on $l^{-1/2}$ (the fit is acceptable), we obtain a finite extrapolated value [dotted line in Fig. \ref{fig:3} (right part)]. On including a term proportional to $l^{-1}$ the visible upward bend of the data is better described. In turn, the extrapolated value of the order parameter increases.
\\ \indent  
For  $J_2=0.2$ a linear fit gives an unphysical value while a quadratic one gives a better description of the data leading, within the accuracy of our calculation, to a null value of the N\'eel order parameter. Therefore it is reasonable to  locate the transition point, at which the N\'eel order vanishes, very close to $J_2 = 0.2$. 
\\ \indent 
For $0.2\lesssim J_2\lesssim0.4$ we find no evidence of magnetic order (either of the N\'eel or collinear type), whereas for larger $J_2$, the GS of the system orders collinearly. 
\begin{figure}[t] 
{\includegraphics[scale=0.68]{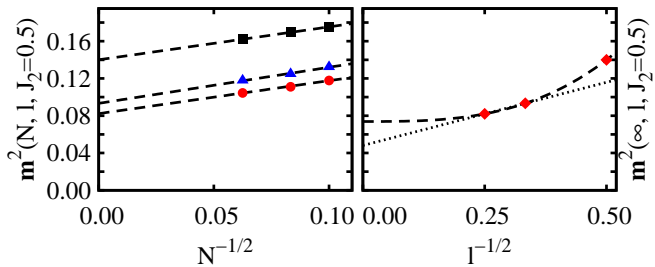}}\\
\caption{(color on line). Same of Fig. \ref{fig:3}  for $J_2=0.5$. Here $A$ and $B$ sublattices are chosen as in the right part of Fig. \ref{fig:1}. The dependence of the collinear order parameter on the plaquette size is described by using a function which includes besides the zero-th order,  orders higher than the quadratic (in $l^{-1/2}$) one (dashed line) and  only the linear  one, excluding the point at $l=4$ (dotted line). Both descriptions yield, when $l$ is saturated, a finite collinear SSM.} 
\label{fig:4}
\end{figure}
\begin{figure}[t] 
{\includegraphics[scale=0.68]{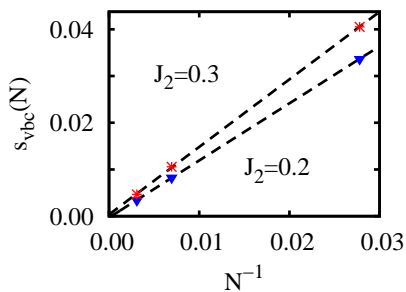}}
\caption{(color on line). Finite size scaling of the plaquette valence-bond crystal order parameter for two values of $J_2$ at which the system is found in a magnetically disordered GS. Dashed lines are fits to the numerical data with the same $J_2$. This order parameter clearly vanishes, for all the $l$ values considered in this work, in the thermodynamic limit. Data shown refer to the $l=16$ case.} 
\label{fig:5}
\end{figure}

Figure  \ref{fig:4} shows the collinear SSM as a function of $N$ (left part), and its values extrapolated to the thermodynamic limit versus $l$ (right part), for $J_2=0.5$. The right  panel shows that the collinear order parameter is finite. Collinear order has been found up to $J_2=1.0$ (i.e., the maximum $J_2$ value considered in this study).\cite{note1} 

Our prediction of N\'eel order persisting up to $J_2\sim 0.2$  is in agreement with  recent  ED works.\cite{honey2, honey3} That the collinearly ordered phase  appears at $J_2\sim 0.4$  is also consistent with the ED suggestion.\cite{honey3} As shown by our energy estimates (Tab. \ref{tab:1}), the collinear  phase is favored with respect to the   dimerized rotational-symmetry breaking  one (described via a resonating valence-bond WF ) proposed for $J_2\gtrsim0.3$ in Ref. \onlinecite{honey1}.   ED studies also predict  a plaquette valence-bond crystal phase in the region in which neither N\'eel nor collinear orders occur. By estimating  the appropriate order parameter, defined as in Ref. \onlinecite{honey2}, we find  this kind of order in the region of interest, only in systems of small size, i.e.,  the  plaquette structure factor is found to vanish in the thermodynamic limit (see Fig. \ref{fig:5}). Therefore, our best GS candidate in this parameter window, is disordered (spin liquid). A  disordered  GS  has been also found by the authors of Ref. \onlinecite{honey1}. Such a phase, however appears at  a value of the NNN coupling constant approximately two times smaller than that found by us. 

Summarizing, the  $J_1-J_2$ model on the honeycomb lattice has been investigated by means of variational calculations based on the EPS  WF. The  phase diagram of the model displays three distinct regions: magnetic order of N\'eel and collinear type is found for ${J_2} \lesssim 0.2$ and ${J_2} \gtrsim 0.4$ respectively; in the intermediate region, none of the ordered parameters considered here remains finite,  consistently with a disordered GS scenario. Such a disordered region has been not revealed by ED studies, in our view, due to the small size of the lattices that are accessible to them.  Interestingly, the value of  ${J_2}$ at which  N\'eel order vanishes, according to both ED and our calculations, is considerably larger than that  ($J_2 \simeq 0.08$) found  in Ref. \onlinecite{honey1}. In Ref. \onlinecite{honey1}, a striking  similarity was suggested between the phase boundary of the $J_1-J_2$ and that of the half-filled Hubbard model \cite{mura} at the Neel-to-disorder transition; the validity of such a suggestion may have to be reconsidered, in the light of the results presented here which indicate that the simple reduction of the physics of the Hubbard model to that of an effective spin-1/2 system, such as the $J_1-J_2$, might not be achievable.

We acknowledge discussions with J. I. Cirac and H.-H. Tu, and thank B. K. Clark for providing us with his energy estimates. This work has been supported by the EU project QUEVADIS, and the Canadian NSERC through the grant G121210893.


\begin{thebibliography}{20}
\bibitem{foot} {Henceforth, we set the value of the NN AF coupling constant $J_1$ to unity, and take it as our energy unit.}
\bibitem{noteA} The large $l$ limit Êof our estimates is achieved by using, starting from plaquettes of a given size (in our case four sites), data obtained with plaquettes whose Êsize is systematically and consistently increased (i.e., by considering plaquettes with $2l^{1/2}+1$ more sites each time that the plaquette size $l$ is changed).
\bibitem{note1} {For $J_2\rightarrow \infty$ the Hamiltonian studied here reduces to that of two decoupled antiferromagnets on the triangular lattice which features   $2\pi/3$ long-range ordered GS, therefore, such a GS has to  emerge when the NNN interaction becomes dominant.\cite{last}}
\end{thebibliography}
\end{document}